\documentclass[twoside,twocolumn]{article}

\usepackage[dvips]{epsfig}
\usepackage[dvips]{graphics}
\usepackage{epsfig}

\usepackage{amsbsy,amsfonts}

\begin{document}

\title{Characterization by a time-frequency method of classical waves propagation in one-dimensional lattice : 
effects of the dispersion and localized nonlinearities.}
\author{O. Richoux, C. Depollier, J. Hardy}

\maketitle


\abstract{This paper presents an application of time-frequency methods to
characterize the dispersion of acoustic waves
travelling in a one-dimensional periodic or disordered lattice made up of
Helmholtz resonators connected to a cylindrical tube. These methods 
allow (1) to evaluate the velocity of the wave energy when the input 
signal is an acoustic pulse ; (2) to display the evolution of the 
spectral content of the transient signal ; (3) to show the role of 
the localized nonlinearities on the propagation \textit{.i.e} the 
emergence of higher harmonics. The main result of this paper is 
that the time-frequency methods point out how the nonlinearities 
break the localization of the waves and/or the filter effects of 
the lattice.}

%
%

\section{Introduction}
When considering propagation of acoustical pulses  in dispersive
media, the crucial question is that of the spreading of the  signals.
Indeed, the investigation of the dispersion relation gives a lot of information about 
the processes which play a part in the wave propagation. Efficient tools for characterizing 
the temporal localization of
the spectral components are the Time-Frequency Representations
(TFR) of signals.
Joint time-frequency representations combine time and
frequency-domain analysis by displaying a signal as a function
defined over the time frequency plane.
 These representations make the behavior of
transient signal visible which is, if not impossible, at least, very difficult
by using  harmonic analysis.
For waves travelling in  strongly dispersive media ({\it e.g.}
a periodic  lattice)  these approaches can be very useful since they allow the
determination of  the dispersion relations 
by measurement of the time of arrival of the different
components (or frequencies)
of  the input signal. When the propagating medium contains  
nonlinearities, the TFR lead also to a detailed
visualization of the nonlinear effects  and show their
consequences for the propagation of signals.

In the following, the dispersion characteristics of sound waves propagating in a waveguide 
with an array of Helmholtz resonators connected axially are examined. From a general model 
of propagation taking account of weakly nonlinear effects localized in the resonators, we 
present Time-Frequency pictures given by acoustic pulses and frequency modulated signals (chirps).

\section{Propagation in a one-dimensional discrete medium}
\subsection{Lattice description}

\begin{figure}[h]
\centering
\psfig{figure=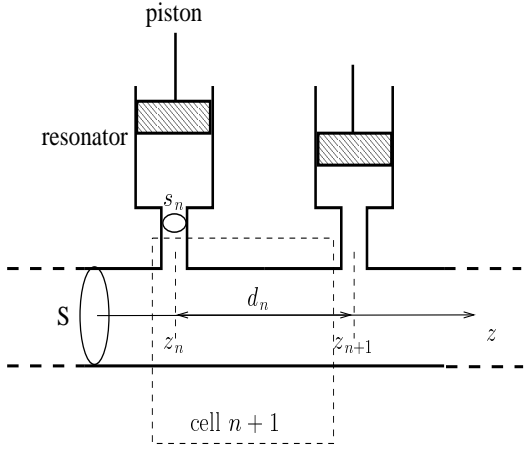,width=7cm,height=6cm}
\caption{\label{fig1}Experimental set up.}
\end{figure}

A one-dimensional lattice made up of
an infinitely long cylindrical waveguide (call also pipe) connected to an array of Helmholtz
resonators numbered by $n$ (simply call resonator hereafter) is considered.
The resonators are connected to the pipe through a pinpoint connection, the radius of the throat's cross sectional area $s_n$ of the $n^{th}$ resonator being assumed to be small compared to the wave length of the acoustic wave ($\sqrt{s_n}/\lambda \ll 1$). 
Each connection is located along the axis of the waveguide by its coordinate $z_n$ with axial spacing $d_n$ for two consecutive points as shown in Fig. (\ref{fig1}).

\subsection{General  case}

In a section of the pipe between two consecutive connection points, the acoustic wave described by the pressure $p(z,t)$ and the acoustic velocity $v(z,t)$ is the solution of the wave equation :
\begin{equation}
\label{eq1}
\frac{\partial^2 p(z,t)}{\partial z^2}-\frac{1}{c^2}\frac{\partial^2 p(z,t)}{\partial t^2}=0
\end{equation}
where $c$ is the sound speed in free space.\\
At each connection point (Fig. (\ref{fig1}), the boundary conditions require the conservation of acoustic flow and continuity of acoustic pressure :
\begin{eqnarray}
\label{eq2a} v(z,t)|_{z_n^+}-v(z,t)|_{z_n^-} & = & -\frac{s_n}{S} v_t(z_n,t) \\
\label{eq2b} p(z,t)|_{z_n^+} & = & p(z,t)|_{z_n^-}
\end{eqnarray}
where $v_t(z_n,t)$ is the acoustic velocity in the throat of the $(n+1)^{\mbox{th}}$ resonator. 
By using the Euler equation, Eq. (\ref{eq2a}) and (\ref{eq2b}) become
\begin{eqnarray}
\label{eq3a} \frac{\partial p(z,t)}{\partial x}\Bigg|_{z_n^+}-\frac{\partial p(z,t)}{\partial x}
\Bigg|_{z_n^-} & = & -\frac{\rho s_n}{S} \frac{\partial v_t(z_n,t)}{\partial t} \\
\label{eq3b} p(z,t)|_{z_n^+} & = & p(z,t)|_{z_n^-}
\end{eqnarray}
where $\rho$ is the air density. They lead to the inhomogeneous wave equation \cite{Levine} : 
\begin{eqnarray}
\label{eq4}
\frac{\partial^2 p(z,t)}{\partial z^2} & - &\frac{1}{c^2}\frac{\partial^2 p(z,t)}{\partial t^2} = 
\nonumber \\
& & \sum_{n}^{} \delta(z-z_n) \frac{-\rho s_n}{S}\frac{\partial v_t(z,t)}{\partial t},
\end{eqnarray}
where the right hand side term has a comb-like structure, the teeth of which having a length related 
to the corresponding impedance jump. This forcing term acts as an array of secondary point-sources 
(scatterers) which work when they are illuminated by the wave travelling in the pipe.

\subsection{Propagation of a monochromatic wave}
\subsubsection{Propagation equation}

\noindent
For a monochromatic acoustic wave with a frequency below the cut-off frequency of the waveguide, 
the acoustic pressure  $p(z,t)$ and velocity $v(z,t)$ along the waveguide are :
$$
p(z,t)=p(z)e^{j \omega t} \quad \mbox{and} \quad v(z,t)=v(z)e^{j \omega t},
$$
where $\omega= kc$ is the angular frequency. The amplitudes $p(z)$ and $v(z)$ are related by an 
impedance relation. At each connection between the waveguide and a resonator, the wave impedance 
(which is modified by a change of cross sectional shape), the
acoustic  velocity  and therefore the derivative of the pressure $dp/dz$, are discontinuous functions. 
Using the Eq. (\ref{eq4}), the pressure $p(z)$ is given by the solution of the following equation : 
\begin{equation}
{\frac{d^2p(z)}{dz^2}}+k^2 p(z)= \sum_{n}^{} \delta(z-z_n)
\sigma_n p(z),
\label{eq:etp}
\end{equation} 
where $\sigma_n= -j\omega \rho s_n/(S Z_n)$. In this relation
$Z_n$ is the impedance of the n$^{th}$ resonator connected at point $z=z_n$ seen from
the guide and $\sigma_n$ is the jump of the pressure derivative $\frac{\partial p}{\partial z}|_{z=z_n}$.

\subsubsection{Matrix method}

\noindent
In the $n+1^{\mbox{th}}$ cell ($z_n \leq z < z_{n+1}$) the pressure and the acoustic velocity are 
respectively denoted by $p_n$ and $v_n$. The solution of the Eq. (\ref{eq:etp}) can be found by the
matrix method \cite{Levine,Dyson,Sugimoto} and the solution $p_n(z)$ is given as a linear combination 
of waves travelling in opposite direction :
\begin{equation}
p_n(z)= A_n e^{jk(z-z_n)}+B_ne^{-jk(z-z_n)}.
\label{eq:sol}
\end{equation} 
Here the coefficients $A_n$ and $B_n$ are respectively
the amplitudes of the forward and backward waves.

\noindent
This function is connected to $p_{n+1}$ in the next region
having passed through the impedance discontinuity at
the connection point $z_{n}$.
The connectivity conditions at the scattering region are the continuity of the pressure 
$$
p_n(z_n) = p_{n+1}(z_n),
$$
and the continuity of the mass flux
$$
{\frac{1}{p_{n+1}(z)}}{\frac{\partial
p_{n+1}(z)}{\partial z}}\Bigg|_{z=z_{n}^+} -  {\frac{1}{p_{n}(z)}}{\frac{
\partial p_{n}(z)}{\partial z}}\Bigg|_{z=z_{n}^-}= \sigma_n.
$$
These junction conditions ensure a physically sound solution
 by requiring continuity all along the waveguide.
Applying the matrix method, these equations
may be reduced to a transfer  matrix ${\cal{D}}_n$  relating the
amplitudes of waves across the junction $z_n$. 
From $z_{n}+\varepsilon$
to  $z_{n+1}-\varepsilon$, the
propagation is modeled by a phase matrix
${\cal{M}}_{n+1}$ 
\[
{\cal{M}}_{n+1}=
\left(\begin{array}{c c}
 e^{jk(z_{n+1}-z_{n})}&
0\\
0&
 e^{-jk(z_{n+1}-z_{n})}\\
\end{array}\right).
\]
The  transfer matrix of
propagation over the region $z_n \leq z < z_{n+1}$ 
 is ${\cal{T}}_{n+1} = {\cal{M}}_{n+1} {\cal{D}}_n $. In this way the
pressure in the duct is given by the following recursive relation : 
\begin{equation}
V_{n+1}=
{\cal{T}}_{n+1}V_{n} \quad \mbox{where} \quad V_n=\left(
\begin{array}{c}
A_n\\
B_n
\end{array}\right) \nonumber
\end{equation}
 which  links the vector pairs   $(A_n \,
B_n)^t$ and $(A_{n+1} \, B_{n+1})^t$. 
The matrix ${\cal{T}}_{n+1}$  has the following form \cite{Levine}
\begin{equation}
{\cal{T}}_{n}=\left(\begin{array}{c c}
\big(1+{\frac{\sigma_{n}}{2jk}}\big) e^{jkd_{n}}&
{\frac{\sigma_{n}}{2jk}} e^{-jkd_{n}}\\
-{\frac{\sigma_{n}}{2jk}} e^{jkd_{n}}&
\big(1-{\frac{\sigma_{n}}{2jk}}\big) e^{-jkd_{n}}\\
\end{array}\right).
\label{eq:ma}
\end{equation}
The propagation through the  lattice  from  $z_n$ to
$z_{n+m}$ is then described  by the relation
\begin{equation}
V_{n+m} = \prod_{i=1}^{m} {\cal{T}}_{n+i} V_n.
\label{eq:ma}
\end{equation}
$V_n$ may be interpreted as  the vector of the initial conditions
(or boundary conditions) and $V_{n+m}$ is the vector of the wave
amplitudes  $m$ cells further along. 

\noindent
The effect of disorder is always to break some symmetry \cite{Ziman}. In 
the case of our lattice disorder breaks the periodicity. So it 
is easier to characterize disordered systems in
terms of their deviations from an ideal of order than it is to define a
perfectly disordered system on which some partial degree of order is to
be imposed,  and,  to think about a disordered system it is
necessary to keep in
mind the ideal system from which it derives.
So, before tackling the problem of wave propagation in  disordered media, 
some specific results about periodic lattices are recalled.

\subsubsection{Periodic lattice}

For an ordered (\textit{i.e.} periodic) lattice, the transfer matrix  
${\cal{T}}_n={\cal{T}}$ is 
the same for every cell, and the infinite medium problem is analogous to
the Kr\"onig-Penney model well known in solid state physics to investigate 
the motion of electrons  in a periodic potential \cite{Dyson}. So 
the propagation in the lattice can be seen in terms of plane waves  subject to
multiple reflections at each derivation, resulting in standing waves.
It is also possible to describe the propagation in terms of a
collective excitation that propagates in the periodic
lattice without scattering but with a modified dispersion relation.
The result is that spatial periodicity gives rise to dispersion
even in the model of plane waves and in this special case the spectrum shows  
frequency intervals (gaps) where no energy propagates.

\noindent
Studying the spectral properties of such a system is then to seek
 stationary states of excitation that satisfy
prescribed conditions at one end of  the lattice. The general 
theory of eigenvalues assures that the
spectrum is not significantly affected by the values chosen as 
boundary conditions, provided the number of cells is large.
The matrix ${\cal{T}}$ has two eiguenvalues $\alpha^{\pm}$ with two
eigenvectors ${\bf W}^{\pm}$ respectively. Any acoustic wave in
the duct can be represented as a linear combination of these eigenvectors
\[
p_n= u^+ {\bf W}^+ +u^- {\bf W}^-.
\]
The effect of operating on $p_n$ by the transfer matrix ${\cal{T}}$
 depends on the nature {\it i.e.} real or  complex of
the eigenvalues $\alpha^{\pm}$. When  they are real, ${\cal{T}}$
simply  "pushes the wave" in the  direction of the eigenvector
corresponding to the greatest eigenvalue. If $\alpha^+>\alpha^-$ then
\[
p_{n+m}= {\cal{T}}^{m} p_n\approx (\alpha^+)^m u^+ {\bf W}^+ 
\]
as $m$ becomes large.  ${\bf W} ^{+}$ is a fixed point or an
invariant point  of the transformation $p_{n+1}=f(p_n)$. By iteration of
this transformation,  ${\bf W}^{+}$ appears as an attractor (or
a sink) in the subspace spanned the eigenvectors : almost all
the waves are attracted by  ${\bf W}^{+}$.
When the eigenvalues are complex, the iteration produces a 
periodic sequence : ${\bf W}^{\pm}$ are indifferent points of the plane.

\noindent
Because the transfer matrix is unitary, the eigenvalues 
of matrix ${\cal{T}}$ obey to the equation
\begin{equation}
\alpha^2+\alpha Tr({\cal{T}})+1=0,
\label{eq:egeq}
\end{equation}
where
\begin{equation}
Tr({\cal{T}})=2\cos{(kd)}+{\frac{\sigma}{k}}\sin{(kd)}=2 \cos(qd)
\label{eq:trace}
\end{equation}
represents the dispersion relation of the Bloch waves \cite{Bloch}, $q$ 
being the Bloch wave number. This  derived dispersion relation exhibits 
the  peculiar characteristic of filters
marked by forbidden frequencies or gaps or stopbands and allowed frequencies or 
passbands in the frequency domain which results from the resonances
and the periodic arrangements of the medium. Waves that obey the relation $|\cos(qd)| \leq 1 $ 
are within a passband and travel freely in the duct and waves  such that $|\cos(qd)| > 1 $ are 
in a forbidden band and are quickly damped spatially. They become evanescent so that they cannot 
propagate. \\
To deduce the particularity of the band structure, Eq. (\ref{eq:trace}) is written in the following 
form
$$
cos(qd)=\frac{\cos (kd+\theta)}{\cos (\theta)}
$$
where $\theta$ is defined by $\tan (\theta)= \frac{-\sigma}{2k}$. The frequencies which bound 
stopbands and passbands are given by the equation :
$$
\cos (kd+\theta)=\pm \cos(\theta)
$$
leading to two kinds of solutions $kd=n\pi$ and $kd=n\pi-2\theta$ with $n \in \mathbb{N}$.

\begin{figure}[h]
\centering
\psfig{figure=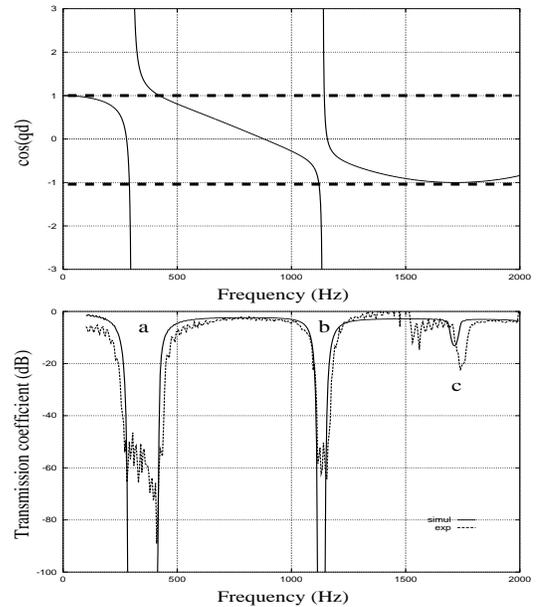,width=7cm,height=8cm}
\caption{\label{fig2}(a) Dispersion relation (from eq. \ref{eq:trace}) of a ordered lattice. 
 (b) Transmission coefficient of the lattice~: $-$ simulation (from eq. \ref{trans-dis}), 
$--$ experiment.}
\end{figure}

\noindent
Two kinds of stopbands appear in the band structure : one kind called Bragg stopband due to the 
periodicity of the lattice, the other due to the resonances of the scatters and called resonance 
stopbands.  A plot of the dispersion relation (\ref{eq:trace}) corresponding to the ordered lattice 
of Helmholtz resonators is showed in Fig. \ref{fig2} which illustrates the different stopbands. 

\subsubsection{The case of a disordered medium}

A disordered lattice is mathematically characterized by a random
sequence of non-identical transfer matrices ${\cal{T}}_n$ whose overall
product describes the propagation along the guide.
\noindent
The disordered linear system gives rise to the classical
Anderson localization \cite{Anderson}.
In this case, all the matrices are different each corresponding to
a different cell, and the wave propagation is described by the
general Eq. (\ref{eq:ma}). The estimation of the Lyapunov exponent, 
from the Ossedelec theorem \cite{Ossedelec}, gives the behavior of the solution 
$p(z)$ in the limit $n\rightarrow + \infty$ :
\begin{equation}
\xi=\lim_{n\rightarrow\infty}{\frac{1}{n}}\ln{(Tr(\prod_{i=1}^n 
{\cal{T}}_i))}.
\label{eq:lya}
\end{equation}
 An alternative way to analyze the
behavior of the waves is the direct evaluation  of  $\tau_{n,m}$
defined, from Eq. (\ref{eq:ma}),  by
\begin{equation}
\label{trans-dis}
\tau_{n,m}=\frac{A_{n+m}}{A_n}
\end{equation}
corresponding  to the transmission coefficient of a lattice made up of $m$ 
cells. As $|m|\rightarrow \infty$, $\tau_{n,m}$ is related to
$\xi$ by  $\tau_{n,m}\approx\exp{(-\xi |m|d)}$
where $d$ is now the mean value of the set of values
$|z_i-z_{i+1}|$ {\it i.e.} the mean length of cells \cite{Luck}.
When the medium is disordered, the gaps widen : eventually
waves cannot propagate in such a medium.
This result is well known since Anderson's original paper \cite{Anderson} : 
in a  one-dimensional disordered 
system, almost all the elementary excitations are localized.
The waves cannot propagate in a disordered medium because
of the many scatterers they encounter. The possibility that a 
wave can be localized in a random medium is mysterious because localization
involves a change in the wave character.

\subsection{Propagation in a nonlinear medium}

For high sound level ($\simeq 120$ dB), the relation between the acoustic pressure 
and the velocity in the neck of the Helmholtz resonators is no longer linear. Nonlinearities 
result mainly from the complicated motion of the air near the tubes between the wave guide and
the neck (throat) of the resonators and at the aperture of each resonator \cite{Rudenko}.\\
Consequently, Eq. (\ref{eq4}) is a nonlinear wave equation where $v_t(z,t)$ 
is not a linear function of $p(z,t)$. Nevertheless we assume that the propagation (between 
two resonators) remains linear. So the pressure in the main pipe is calculated in the same 
way as for a disordered medium but now nonlinear operators describe the scattering of waves at each 
connection point of the lattice. 
A simple model of nonlinear Helmholtz resonator is developed in the following by using a Taylor's 
development of the restoring force due to the change of pressure in the Helmholtz cavity. 

\section{A simple nonlinear Helmholtz resonator model}

\noindent
It is well known that a simple model of the Helmholtz resonator requires 
the following assumptions : (1) the pressure inside the cavity is 
spatially uniform, (2) the fluid in the neck moves like a solid piston. 
In this case, the air enclosed in the resonator acts as a spring for the 
lumped mass of air moving within the neck. A general description of 
the nonlinear 
behavior of the Helmholtz resonator may be derived by taking into 
account the quadratic term in the restoring force of the spring. 
For thermodynamical processes that occur in the air within the cavity, 
the adiabatic changes of the pressure $p$ and the volume $V$ are related 
by $p_0/p=(V/V_0)^{\gamma}$ where the subscript $0$ stands for the 
unperturbed reference values and $\gamma$ is the specific heat ratio. 
The relative change of the pressure $p_n$ in the cavity of the 
n$^{th}$ resonator due to a 
small displacement $x_n$ of the air in the neck induces a restoring 
force $F_n$ that has the following form \cite{Boullosa} : 
$$
F_n = p_n s_n = - \frac{\rho c^2 s_n^2}{V_0}[x_n - \alpha_n x_n^2 + o(x_n^3)].
$$
In this equation $\alpha_n = (\gamma+1) s_n / (2 V_0)$ and $c$ is the 
sound velocity given by $c=\sqrt{\gamma p_0/ \rho}$. 
The spring force is no longer linear and its stiffness is now described by 
two constants. For a monochromatic wave the displacement $x_n$ of the air in the neck is related 
with the acoustic velocity $v_n=v(z_n)$ by the relation $v_n=j \omega x_n$ and the Euler relation 
applied to the air mass $m=\rho l'_c s_n$ (where $l'_c$ is effective neck length) 
submitted to the harmonic force $p_n/(\rho l'_c) e^{j \omega t}$ gives :
$$
j \omega v_n + \omega_0^2 [\frac{v_n}{j \omega}-\alpha_n ( \frac{v_n}{j \omega} )^2 + o(v_n^3)]
=\frac{p_n}{\rho l'_c}
$$
where $\omega_0^2=s_n c^2 /(V_0 l'_c)$ is the resonance frequency of the Helmholtz resonator.
The nonlinear relation between the acoustic pressure and 
the velocity just outside of the opening of the n$^{th}$ resonator is 
$$
p_n=Z^{NL}_n v_n.
$$
Here $Z^{NL}_n$ is the "nonlinear impedance" of the n$^{th}$ resonator : 
$$
Z^{NL}_n=Z^L_n + \alpha_n \rho l'_c \frac{\omega_0^2}{\omega^2} 
\frac{p_n}{Z^L_n} + o\left( (\frac{p_n}{Z_n^L})^2 \right),
$$
and $Z^L_n$ is the linear impedance of 
the n$^{th}$ resonator ($Z^L_n=j \omega \rho l'_c(1-
\frac{\omega_0^2}{\omega^2})$). So, the nonlinear effects lead 
merely to a additive correction to the linear impedance of the Helmholtz 
resonator which is non-vanishing only around the Helmholtz resonance $\omega_0$. 
Others nonlinear phenomenon can occur due to the presence of turbulences (vortex) 
which are generated around the edges of the neck \cite{Rudenko}. According to the shape 
(sharp or rounded) of these edges, the physical phenomena lead to nonlinear terms more 
or less important \cite{Dequand}.

\section{Signal analysis}

\noindent
Time-frequency representations have several advantages over 
conventional harmonic Fourier analysis. Among these one can cite their
ability to analyze multicomponent signals {\it i.e.} 
broadband signals and  also signals containing 
several modes which interfere or combine each other.  
But the most important advantage is their capacity to track 
the time evolution of transient signals.
This is exemplified by the visualization  of reflected signals
due to scattering waves and of the spreading of acoustic pulses
in the joint time frequency plane.
In this paper the quadratic TFR (QTFR) are used 
to determine empirically the dispersion of the waves since this
provides a direct access to the group velocity and allows to
extract pertinent information associated with the waves
travelling in the lattice.
Unfortunately, QTFR also have some
drawbacks. As it is  bilinear in the signal, the QTFR of the sum
of two signals is not the sum of the QTFR of each of them but 
has an additional cross term which gives rise to
artefacts and confuses the picture in the time-frequency plane.
Several solutions are
available for reducing these terms, for example, by using  an
appropriate  smoothing
window function and/or the analytic signal.

\noindent
For the complex-valued  time signal $x(t)$, the general form of the QTFR is \cite{Cohen}
\begin{eqnarray}
W_x(t,\omega) & = & {\frac{1}{2\pi}}\int\int\int
e^{-j\vartheta t-j\tau\omega-j\vartheta u}\phi(\vartheta,\tau) \nonumber \\
& & x\big(t+{\frac{\tau}{2}}\big)
x^*\big(t-{\frac{\tau}{2}}\big)  d\tau du d\vartheta
\label{eq:tfr}
\end{eqnarray}
where $\phi(\vartheta,\tau)$ is the kernel of the QTFR, $^*$
stands for complex conjugate and $t$
and $\omega$ are time and frequency respectively. 
The most popular QTFR is the Wigner-Ville distribution \cite{Meckenlenbrauker} for which
the kernel is $\phi(\vartheta,\tau)=1$ :
\begin{equation}
W_z(t,f)=\int_{-\infty}^{+\infty} z\big(t+{\frac{\tau}{2}}\big)
z^*\big(t-{\frac{\tau}{2}}\big) e^{-2j\pi f \tau} d\tau.
\label{eq:wv}
\end{equation}
To reduce the artefacts arising from the cross terms between the different components of 
the signal, we use the analytic signal \cite{Gabor}
$z(t)$ defined by
\begin{equation}
z(t)=x(t)+j H[x(t)]
\end{equation}
where $H[x(t)]=\frac{1}{\pi t} \star x(t)$ is the Hilbert transform of $x(t)$ designed by the 
Kaiser window method (FIR Hilbert transformer) \cite{Oppenheim}. The Pseudo Wigner-Ville distribution 
is given by  
\begin{eqnarray}
PW_z(t,f) & = &\int_{-\infty}^{+\infty} \vert h(\tau)\vert^2 \nonumber \\
& &z\big(t+{\frac{\tau}{2}}\big)
z^*\big(t-{\frac{\tau}{2}}\big) e^{-2j\pi f \tau} d\tau,
\label{eq:pwv}
\end{eqnarray}
where $h(\tau)$ is a smoothing  window (in our case a Kaiser function).

\section{Experimental results}
\subsection{Experimental apparatus}

\noindent
Figure \ref{fig1} shows the experimental setup.
It consists of a 8 m long cylindrical  pipe having a 5 cm  
inner diameter and a 0.5 cm thick wall
connecting with an array of  60  Helmholtz 
resonators as side branches. The distance between two consecutive resonators is $d_n=0.1$ m. 
The upstream section links this system to a loudspeaker designed for high acoustic power level and 
used to generate linear frequency modulated waves (chirps) or wavepackets (approximate $\delta$-function). 
The duration of the shirp is $0.5$ s and the frequency range is included in $[100;600]$ Hz. The wavepackets 
duration is 0.01 s and it frequency range extends from $0$ to $2500$ Hz. At the  end of the downstream 
section, an approximately anecho\"ic termination made of plastic foam suppresses reflected waves. 
As noted above, the QTFRs are able to discriminate the
reflected signal from the incident one and an anecho\"ic
termination seems unnecessary. However, its use prevents a too
heavy contamination of the useful picture in the
time-frequency plane  by unessential signals and thus improves the
accuracy of the QTFR. Lastly, two microphones $m_1$ and $m_2$ (B\&K $4136$ with 
Nexus $2690$ amplifier)
measure the pressure 
in up and downstream sections. These microphones produce $0.2$ \% of distortion at $150$ dB which 
ensures that the nonlinear effects are generated by the propagation under the lattice (and not by 
the microphones themself). The data acquisition is carried out by means of a $16$ bits AD-converted 
with a sampling frequency of $10$ kHz and an anti-aliasing filter ($8^{th}$ order Chebyschev filter) 
with a bandwidth of $4$ kHz .

All the  resonators are identical and are $L=16.5$ cm long cylindrical cavities with a diameter of $4$ cm. 
Their volumes may be independently tuned by moving pistons.  Their neck is a $2$ cm long tube having a 
diameter of $1$ cm. The  Helmholtz resonance frequency corresponding to the whole 
volume ($V_0=2.1 \, 10^{-4}$ m$^3$) is $300$ Hz. All  the pipes  used are  sufficiently stiff that their 
structural modes are not in the frequency range of interest.
  
\subsection{Linear case}

\begin{figure}[h]
\centering
\psfig{figure = 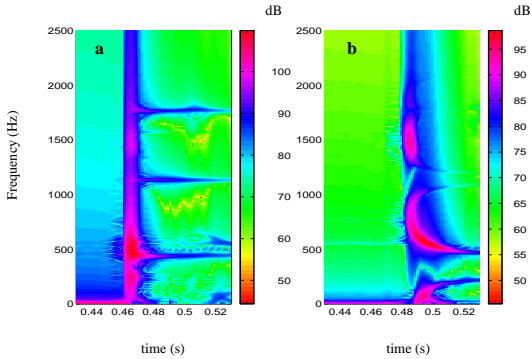,width=7cm}
\caption{\label{fig3}(a)  Wigner-Ville transform of a pulse signal upstream of the lattice.  
(b) Wigner-Ville transform of pulse signal downstream of the lattice. }
\end{figure}

\noindent
Figure \ref{fig2} gives the theoretical [$f$, $\cos{(qd)}$] plot of the
dispersion relation of Bloch waves for the ordered lattice
described above (Fig. \ref{fig2}-a) and the experimental results
deduced from the transfer function of the system (Fig. \ref{fig2}-b). We
observe a good fit in the frequency range of interest. For
this geometry the curve shows two kinds of gaps : those (labeled a and b) due to the resonators 
included in $[300:450]$ Hz (Helmholtz resonance) and $[1100:1200]$ Hz (corresponding to the length 
of the cavity $L$) and one other that is a specific characteristic of the periodic lattice ($d_n=0.1$ m) 
at $1700$ Hz (labeled c).
\noindent
The Pseudo Wigner-Ville Distribution images associated with the
propagation of an acoustic pulse in the lattice are presented in 
Fig. \ref{fig3}.
Figure \ref{fig3}-a is the PWVD of the signal picked-up in the upstream
section of the apparatus by microphone $m_1$ corresponding to the incident impulse beginning at $0.4$ s 
and three narrow bands reflected signal. Fig. \ref{fig3}-b shows 
the PWVD of the transmitted (output) signal given by microphone $m_2$ (no reflection are visible because 
of the anechoïc termination). As well as the time frequency plot of the incident impulse, other
details can be seen in Fig. \ref{fig3}-a. For time $t>0.47$ s a part of
the input signal is reflected corresponding to the upper cutoff frequencies 
of the gaps. The two lower stopbands correspond to the energy of acoustic waves reflected by the 
resonators. 

The third case has a different nature since this gap is due to the 
spatial periodicity of 
the lattice. The reflected signal corresponding to this stopband 
indicates that at least for 
a finite length lattice, the spectrum structure 
(gaps and allowed bands) is not produced 
only by wave interferences but also by some wave
reflection. This result cannot be displayed by a method other than
QTFRs since it requires time and frequency to be localized. Some authors 
have investigated analyticaly \cite{Sugimoto} or experimentaly \cite{Bradley} the dispersion 
characteristics of sound waves in 
a lattice of Helmholtz resonators but they have not pointed out the 
existence of a reflected wave for particular bands.
 
\begin{figure}[h]
\centering
\psfig{figure = 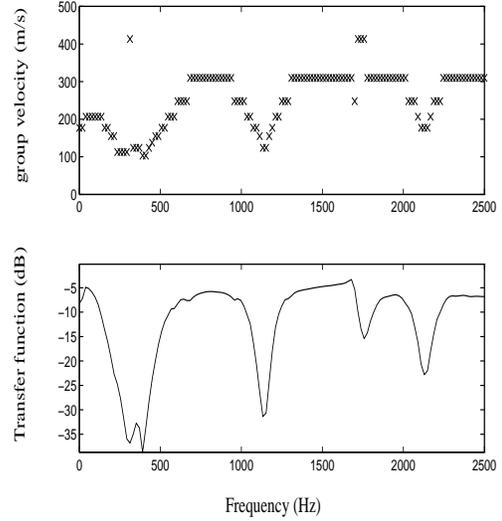, width=7cm,height=7cm}
\caption{\label{fig4}Group velocity vs. frequency and transfer function of the lattice.}
\end{figure}

\noindent
In Fig. \ref{fig3}-b, the waves belonging to the allowed bands are shown
at the time $0.49$ s corresponding to the propagation
of waves through the lattice. The gaps are easily identified 
and are in good agreement with the values given by the Bloch theory.
The waves with frequencies closed to the gaps arrive at longer and
longer delays : their group velocity goes to zero as the frequency
gets closer to the cutoff frequencies. This is indicated on the PWVD
plot by long tails on each side of the gaps. From the point coordinates
of PWVD of the output signal, it is possible to estimate the group 
velocity as a frequency dependant function (Fig. \ref{fig4}).  

\subsection{Nonlinear case}

\begin{figure}[h]
\centering
\psfig{figure = 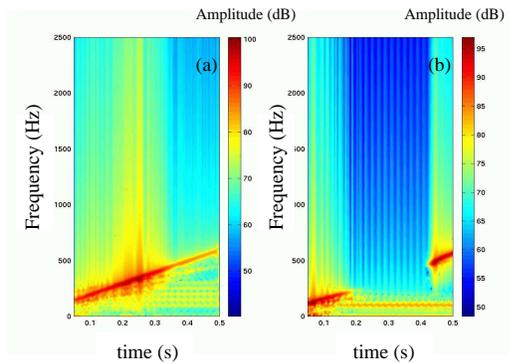, width=7cm}
\caption{\label{fig5} (a)  Wigner-Ville transform of a FM signal upstream of the lattice 
(linear case). (b) Wigner-Ville transform of a FM signal downstream of the lattice (linear case).}
\end{figure}

\begin{figure}[h]
\centering
\psfig{figure = 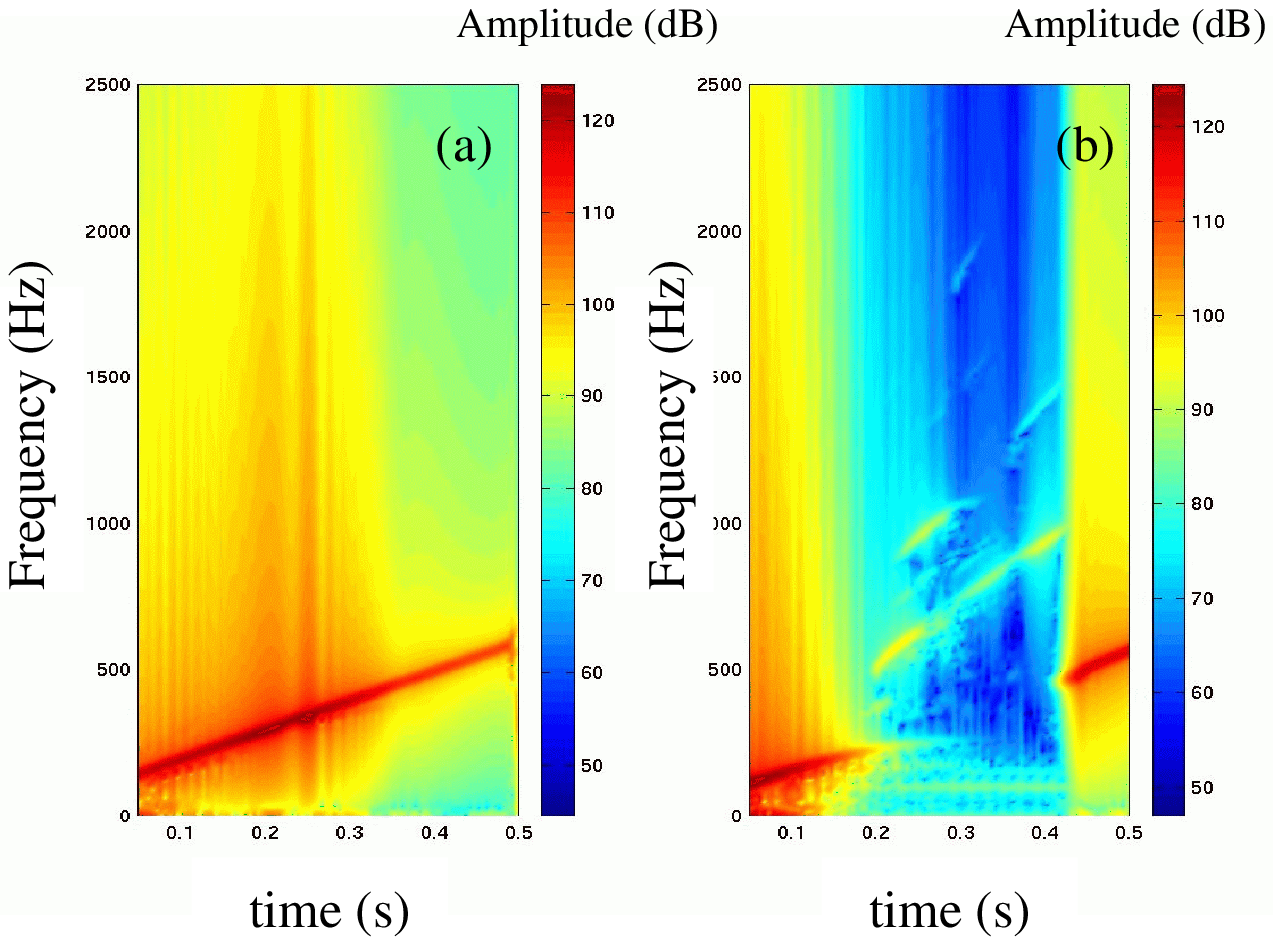, width=7cm}
\caption{\label{fig6}(a) Wigner-Ville transform of a FM signal upstream of the lattice 
(nonlinear case). (b) Wigner-Ville transform of a FM signal downstream of the lattice (nonlinear case).}
\end{figure}

\begin{figure}[h]
\centering
\psfig{figure = 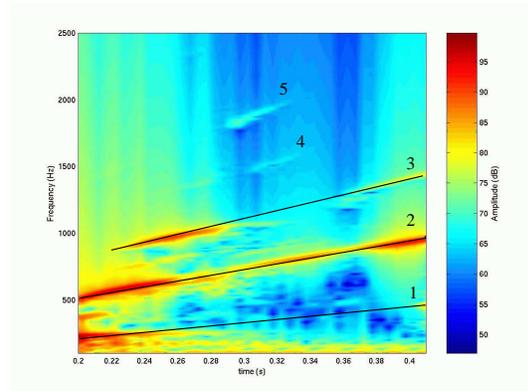, width=7cm}
\caption{\label{fig7}Zoom of the figure (\ref{fig6}b) in the time range $[0.2:0.42]$ s. Numbers label 
the corresponding harmonics from the fundamental ($1$) to harmonic $5$.}
\end{figure}

\noindent
In the nonlinear case, the physical parameters of the output
are dependent on the input parameters. It is then necessary to 
consider several cases with
different input conditions. To investigate 
the effects of these nonlinearities, the input signal is a 
linear FM signal with a frequency range including the first 
stopband of the lattice (from $100$ Hz to $600$ Hz). Fig. \ref{fig5} and Fig. \ref{fig6} 
show the PWVD of this chirp for low and high acoustic levels  
corresponding to the linear and the nonlinear cases respectively. 
In the linear case, waves belonging to the first forbidden band cannot propagate, as shown by 
Fig. \ref{fig5}-b. For higher acoustic levels, nonlinearities generate 
harmonics $2 \omega$ and $ 3 \omega$ which belong to allowed bands (Fig. (\ref{fig7}) which represents 
a zoom in the time scale range $[0.2s :0.4s]$ of the Fig. (\ref{fig6})). Then, 
for some particular frequencies in the first stopband the nonlinearities 
take part to the delocalization of the energy and gives rise to energy transport : the energy of an input 
signal contained in a frequency range which is in a gap propagates in the 
lattice after being transferred 
in ranges corresponding to allowed bands in the spectrum. This mechanism breaks the filter structure 
of the lattice and is partly responsible for changes in the transmissivity 
of the medium. Indeed, the amplitude is restored with presence of 
nonlinearity : each site exhibits a nonlinearity and for localization 
lengths of the order of the lattice spacing, the associated wave becomes 
propagating. The measurement of the energy amplitude of each harmonics can permit the estimation of 
coefficients $\alpha_n$ (and the next in  the Taylor's development) corresponding to the harmonics in 
the model developed in section 3. The Fig. (\ref{fig7}) shows also the different phenomena induce by the 
nonlinear effects : in the beginning of the stopband (for time including between $0.2$ and $0.25$ s) only 
harmonic $2$ is present in the response of the lattice whereas in the rest of the stopband harmonics $3$, 
$4$ and $5$ are detected in the time-frequency image. Because the nonlinear effects are not the same in 
all the stopband, we presume that the different coefficients of the nonlinearities in the model developed 
in section 3 are depending on the frequency of the wave : these effects result from the competition between the different orders of the nonlinearities \cite{Richoux2}. A experimental study of one Helmholtz resonator confirms this assumption and allows the measurement of the different coefficients of the model \cite{Richoux}.

\section{Conclusion}

In this paper we have shown that the time-frequency methods are
well adapted to the investigation of the propagation of
dispersive waves. Especially they give information about
the group velocity and allow to find the dispersion characteristics
of complicated waveguide. These methods are also very efficient
for displaying the  nonlinear behavior of high-intensity  sound
waves : the generation of higher harmonics can be directly measured.
Lastly, the interpretation of experimental results of wave
propagation in nonlinear lattices is made easier
with  the help of time-frequency methods.


\end{document}